\documentclass[preprint]{aastex}

\slugcomment{accepted for publication in PASP, Jan 2004}

\shorttitle{Minimising Telluric absorption in NIR spectra}
\shortauthors{Kenworthy \& Hanson}

\begin{document}

\title{Minimizing Strong Telluric Absorption in\\
Near Infra-red Stellar Spectra}

%% Use \author, \affil, and the \and command to format
%% author and affiliation information.
%% Note that \email has replaced the old \authoremail command
%% from AASTeX v4.0. You can use \email to mark an email address
%% anywhere in the paper, not just in the front matter.
%% As in the title, you can use \\ to force line breaks.

\author{Matthew A. Kenworthy\altaffilmark{1} and Margaret M. Hanson}
\affil{Department of Physics, University of Cincinnati, Cincinnati OH
45220}
\email{matt@physics.uc.edu}

\altaffiltext{1}{Steward Observatory, 933 North Cherry Avenue, Tucson, AZ 85721}

\begin{abstract}

We have obtained high resolution spectra ($R \sim 25000$) of an A star
over varying airmass to determine the effectiveness of telluric removal
in the limit of high signal to noise.  The near infra-red line
\ion{He}{1} at $2.058\micron$, which is a sensitive indicator of
physical conditions in massive stars, supergiants, \ion{H}{2} regions
and YSOs, resides among pressure broadened telluric absorption from
CO$_2$ and water vapor that varies both in time and with observed
airmass.

%The time-varying nature of the atmospheric absorption requires careful
%observations of standards interleaved with the science targets, and so
%we investigate the variability of the telluric absorption at high
%spectral resolution ($\rm{R}\sim 24,000$) by monitoring a standard star
%through a range of airmasses under typical observing conditions over
%$2.0584-2.0636\micron$.

Our study shows that in the limit of bright stars at high resolution,
accuracies of 5\% are typical for high airmass observations (greater
than 1.9), improving to a photon-limited accuracy of 2\% at smaller
airmasses (less than 1.15).  We find that by using the continuum between
telluric absorption lines of a ro-vibrational fan a photon-limited 1\%
accuracy is achievable.

\end{abstract}


\section{Introduction}

Our goal is to obtain high quality moderate resolution spectra ($R \sim
10000-20000$) to be used in conjunction with sophisticated stellar model
atmospheres \citep{san97}.  The near infra-red (NIR) spectrum of the
atmosphere from 1 -
5 $\mu$m, is dominated with telluric water and CO$_2$ absorption,
between which the major observing windows in the wavelength region are
defined. Broad molecular absorption bands of CO$_2$ are resolved at
moderate spectral resolution (R $>$ 5000) into individual pressure
broadened ro-vibrational transitions.  These transitions vary both in
time and observed elevation making their removal problematic in ground
based spectroscopic studies.

There is a tremendous motivation from astronomers to obtain highly
accurate spectral profiles for lines existing in the near-infrared.
For hot stars, an important line in modeling extended atmospheres
and stellar winds is the \ion{He}{1} transition at $2.058\micron$
\citep{han96}. This line is also an important diagnostic
in \ion{H}{2} regions \citep{lum03}, massive
young stellar objects \citep{port98} and planetary
nebulae \citep{ds94,lum01}.  However, the line is located near
the short wavelength edge of the K band window, where CO$_2$ absorption
bands dominate the spectrum.  The challenge for observational
astronomers is in observing the $2.058\micron$ line profile with high
signal to noise suitable for detailed modeling despite considerable
telluric contamination.

Observations of standard stars are essential to obtain even modest
correction of telluric absorption features in this spectral region. This
paper investigates the variability of the telluric absorption near the
\ion{He}{1} transition and discusses the limits to the signal to noise
ratio achievable given a typical observing program and instrument. We follow
an A2V star for half a night through a range of airmasses and see what
effect the variation of telluric absorption has on the recovered
spectra.

\section{Observations}

\begin{figure}
\includegraphics[angle=0,width=\columnwidth]{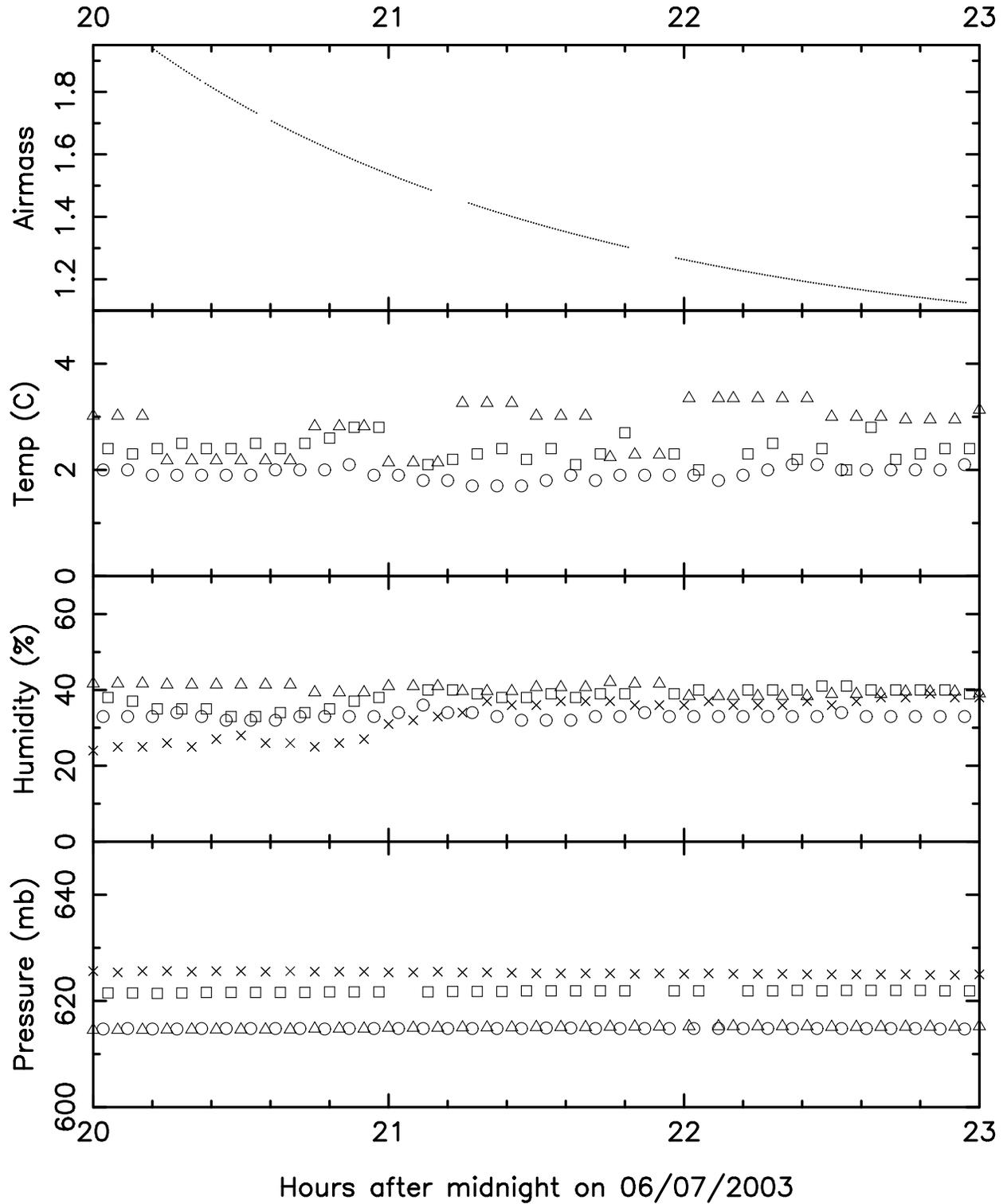}
\caption{Meteorological conditions from different telescope weather
stations across the summit of Mauna Kea. CFHT data are marked with a
circle, CSO with a cross, Subaru with a square, and UKIRT with a
triangle. Data provided by Mauna Kea Weather Center (Lyman 2003, priv.
communication).
\label{weather}}

\end{figure}

The data were taken at the IRTF on 2003 June 08 0600 UT (2003 June 07
2000 HST).  Meteorological data from weather stations across the Mauna
Kea summit are shown in Figure \ref{weather} along with the range of
airmasses observed. The observations started at an airmass of 1.95 and
finished at an airmass of 1.13, when the star was approaching transit.

We use the CSHELL echelle spectrograph \citep{gre93} with a slit width
of 1.0 arcsec.  Combined with the pixel scale of 0.2 arcsec/pixel, this
gives a slit width of 5 pixels on the $256\times256$ ($30\micron$ pitch)
InSb array.  The array has a read noise of 55$e^-$ per pixel per sample,
and a dark current of 0.5 $e^-$ per second per pixel with an approximate
gain of 11 $e^-/$ADU.  With a 325mV bias, it is 1\% linear at
22,000$e^-$. One echelle order is selected with an order sorting filter,
and the dispersion is 0.20\AA /pixel. Due to problems with the telescope
guiding camera, the telescope is off-axis guided by a nearby star, in
contrast to the usual mode of on-axis guiding with the target star in
the visible.

The dispersion of the spectrograph at $2.0563\micron$ is measured at
0.207\AA/pixel. The instrumental spectral resolving power is measured by
fitting a Gaussian to two OH night sky emission lines, identified as a
$\Lambda$ P1 doublet \citep{rou00}, with a full width half maximum
(FWHM) of 4.0 pixels ($12.1$km/s) with the star overfilling the slit.
This gives a higher spectral resolution than expected with the stated
slit width, and so we take our measured value of the dispersion and
instrumental width to obtain a spectral resolution of 24,800.

%% 20562.953\AA and 20564.143\AA 

A dwarf A2 star is not expected to have any significant metal lines
in its spectrum, and as a result acts as a continuum source above the
atmosphere over the spectral region of the study. The A2V star
\objectname[HD 156729]{HD 156729} $(17\fh17\fm40.2545\fs,
+37\fdg17\farcm29.39\farcs, J2000, V=4.62)$ is followed from an airmass of
1.95 through to 1.13 with 30 second integrations in a continuous beam
switching mode with a throw of $9.0\arcsec$. The mean time between the
start of successive observations is 34 seconds, resulting in 268 fully
reduced observations.

We set the central wavelength of the spectra to $2.061\micron$ to
include a range of different line strengths. Three P branch transitions,
the Q transition, and four R branch transitions of CO$_2$ are included
in the spectra, along with some H$_2$O transitions (see Fig.\ 2).

A set of twenty flat fields is taken immediately after the end of the
observations, using a continuum lamp placed in front of the entrance
slit of the spectrograph. Dark frames of identical duration to the flat
field exposures are taken later in the night.

\section{Reduction of the Echelle Spectra}

All spectra are reduced using \tt IRAF\rm \footnote{IRAF is distributed
by the National Optical Astronomy Observatories, which are operated by
the Association of Universities for Research in Astronomy, Inc., under
cooperative agreement with the National Science Foundation.} routines,
and subsequent analysis is done using PerlDL\footnote{homepage at:
\url{http://pdl.perl.org/} }.

The dark frames and flat field frames are averaged together to form a
master dark frame and master flat frame and frames showing the r.m.s.
per combined pixel (sigma frames) are produced. The master dark frame is
subtracted from the master flat field frame, and the flat field is
normalized in the spectral direction to produce a pixel response frame.
Bright, noisy and dead pixels are identified using the sigma frames,
and a bad pixel frame is generated for the detector. Pairs of science
frames are subtracted from each other, divided by the pixel response
frame and then bad pixels flagged in the bad pixel frame are
interpolated over using data from the surrounding good pixels.

There are two spectra per beam-switched frame. A fixed aperture width of
24 pixels (corresponding to $4.8\arcsec$) is used to extract the spectra
from the frame. Although in ideal beam switching the frame background
level is zero, we noted in four pairs of beam-switched frames that a jump
in bias level in the readout of the frame would result in a zero level offset
of up to 20 counts being present in the data, compared with the typical
flux of 200 counts per pixel in the spectrum. Visual examination of
these images show a very smooth background that is automatically
subtracted as part of the spectral extraction. Remaining transient
events are removed using the 'clean' algorithm in the spectral
extraction package, using the known gain and effective read noise of the
beam switched images. The centroid position of the spectra are traced
with a third order Legendre function.

\begin{figure}
\includegraphics[angle=270,width=\columnwidth]{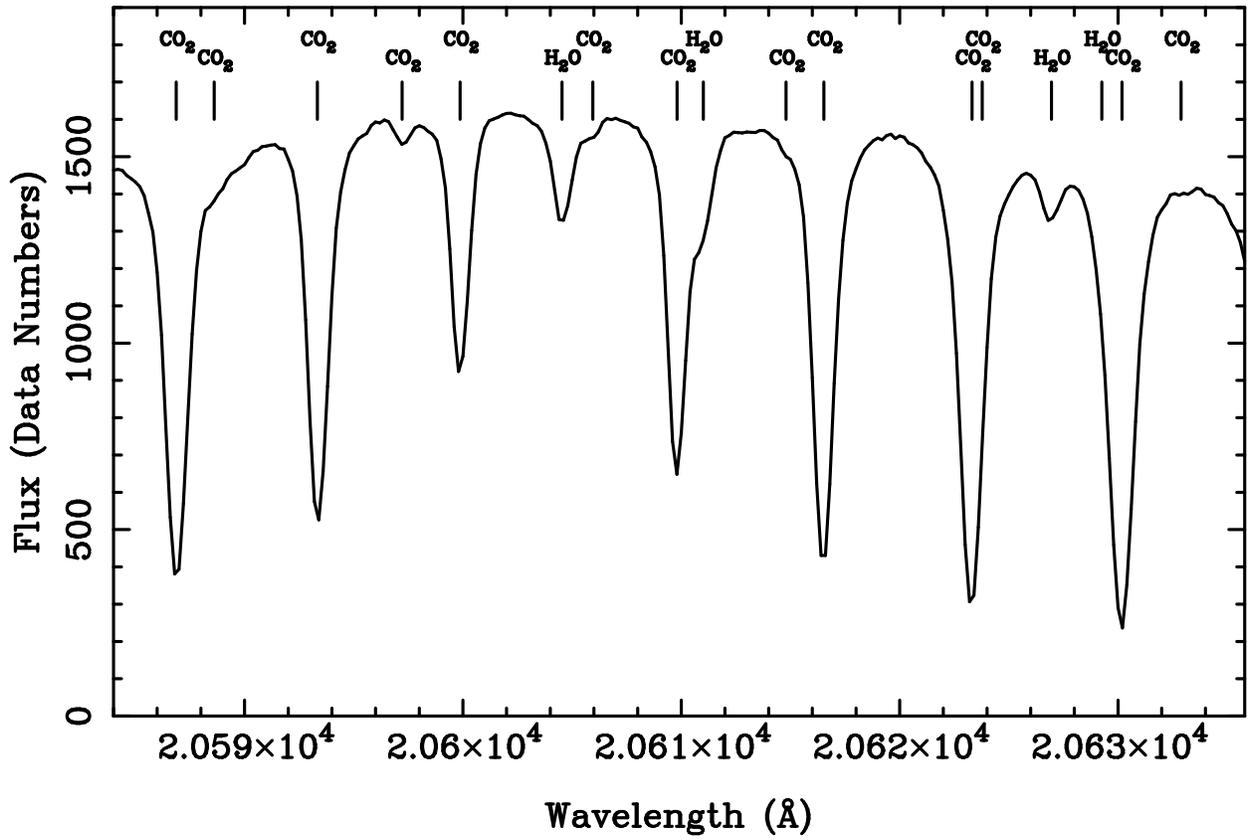}

\caption{Spectrum of an A2V star, averaged from 10 exposures at a mean
airmass of 1.13. All spectral features are due to telluric absorption in
the Earth's atmosphere.  Lines are identified according to the HITRAN database
\citep{rot03}. Echelle transmission function has not been removed from
the spectrum.  \label{spectrum}}

\end{figure}

Wavelength dispersion correction is performed using the telluric
profiles themselves. A model of the atmospheric transmission is
generated for the observing site using ATRAN \citep{lor92} and all
absorption line profiles are measured from this reference. These are
then used to wavelength calibrate the individual spectra. Using a second
order Legendre function, a wavelength solution with rms fit of 0.03\AA~
is typical, corresponding to a goodness of fit of about 1/10 of a pixel.
The absorption lines are labeled according to the HITRAN database
\citep{rot03} in Figure \ref{spectrum}.

\section{Observing Technique}

The telluric absorption in the NIR varies as a function of airmass and
of time \citep{tok99}. Ideally, in spectral regions of strong telluric
absorption, a standard star and a science target should be observed
through identical atmospheric paths at the same time, allowing the
atmospheric absorption to be divided out. In this perfect case, a
reference star would be within a few arcseconds of a science target, and
placing the spectrograph slit across both objects in a position suitable
for beam-switching would result in the highest observing efficiency and
accuracy.

Bright stars with no metallic lines (usually those later than B8V and
earlier than A2V), are the most appropriate for telluric removal. They
are not perfect in that these stars have broad Hydrogen lines in their
spectra and to obtain science target spectra in these regions, various
modeling techniques have been developed \citep{mai96,han96,vac03} to remove
the Hydrogen lines from the standard star spectra whilst preserving the
(much narrower) telluric absorption lines.

\subsection{Standard star observations at different airmasses}

\begin{figure}
\includegraphics[angle=270,width=\columnwidth]{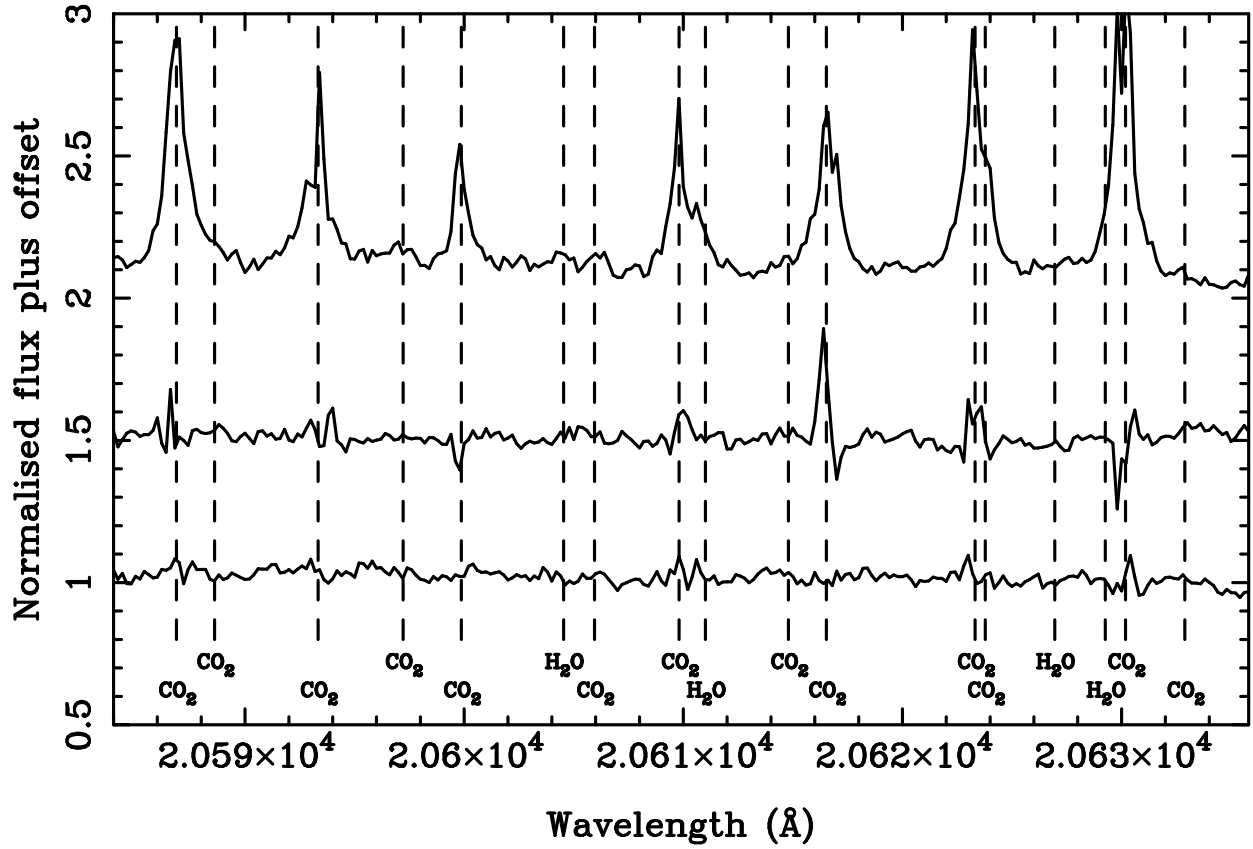}

\caption{The ratio of two spectra of the same star taken at different
airmasses. The lowest spectrum is the division of two spectra taken at
an airmass of 1.13. The middle spectrum is the division of spectra taken
at an airmass of 1.94, and the top spectrum shows the division of a
spectrum at an airmass of 1.13 by a spectrum at 1.94. Each spectrum has
been offset by a flux of 0.5, and the dashed lines mark the center of
telluric absorption lines and their associated molecular species.
\label{twodivs}}

\end{figure}

A complete observation consists of spectra of the science target,
slewing the telescope to a nearby standard star, and taking spectra of
the standard star. The standard star and science target observations are
separated both in time and position on the sky, and so are observed
through different paths in the atmosphere. Differences in the telluric
absorption between these two pointings result in systematic errors
appearing in the science object spectra, and these can significantly
reduce the signal to noise and sensitivity reached in the object
spectrum. This is shown in Figure \ref{twodivs}, where the ratios of
standard star spectra at 1.13 and 1.94 airmasses is shown. The two lower
spectra show the ratio of spectra taken approximately 35 seconds apart,
with the lowest spectrum at 1.13 airmasses and the middle spectrum at
1.94 airmasses.  The spectra show a flat, Gaussian noise dominated
continuum, with spike-like residuals located at the cores of telluric
absorption lines. This effect is seen in greater detail in the topmost
spectrum, where the ratio is between a spectrum from 1.13 and 1.94
airmass.

The root mean squared value of such normalized spectra is a measure of
the standard deviation of the spectrum from its mean value.  Various
sources of noise combine together in quadrature to give the theoretical
lowest noise limit attainable for a given spectral element in the final
spectrum. For our ratioed spectra, we calculate a theoretical signal to
noise ratio and compare it to the measured signal to noise ratio. The
noise sources we include in the calculation of the theoretical noise
limit are detector read noise, photon count noise, and flat field noise.
For our extraction width of 24 pixels this corresponds to a read noise
of 380$e^-$ per wavelength bin of a beam-switched image, and our flat
field signal to noise ratio is approximately 1200.

\begin{figure}
\includegraphics[angle=270,width=\columnwidth]{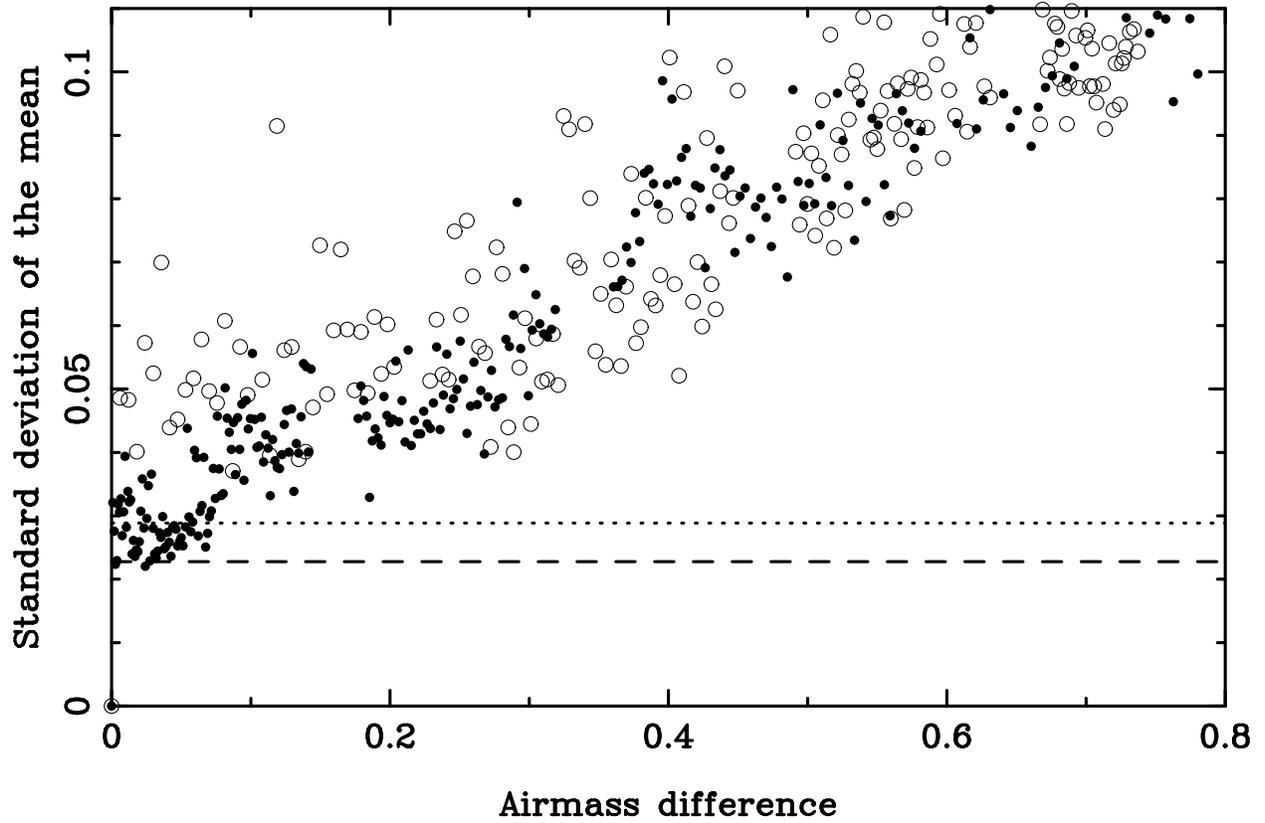}

\caption{A graph showing the standard deviation of the mean of the ratio
of a standard star taken at two different airmasses, over a wavelength
range of $2.0584-2.0636\micron$. Filled circles relative to an airmass
of 1.13, hollow circles relative to an airmass of 1.94. The dashed line
represents the theoretical signal to noise attainable for observations
at an airmass of 1.13, the dotted line is the same but for an airmass of
1.94.  \label{amdiff}}

\end{figure}

The increase in airmass difference between the two spectra leads to
lower signal to noise, as seen in Figure \ref{amdiff}. At low airmass,
the theoretical limit of 2.3\% is reached for exposures taken one minute
apart (the filled circles).  However, at a higher airmass of 1.94
(hollow circles), its theoretical limit of 2.6\% is NOT reached, and
instead the noise tails off at 5\%, a signal to noise ratio of only 20.
Seeing the significant telluric residuals for 1.94 airmasses in the
middle spectrum of Figure \ref{twodivs}, it is apparent that as low
as possible an airmass should be used for both standard and
science object.

\begin{figure}
\includegraphics[angle=270,width=\columnwidth]{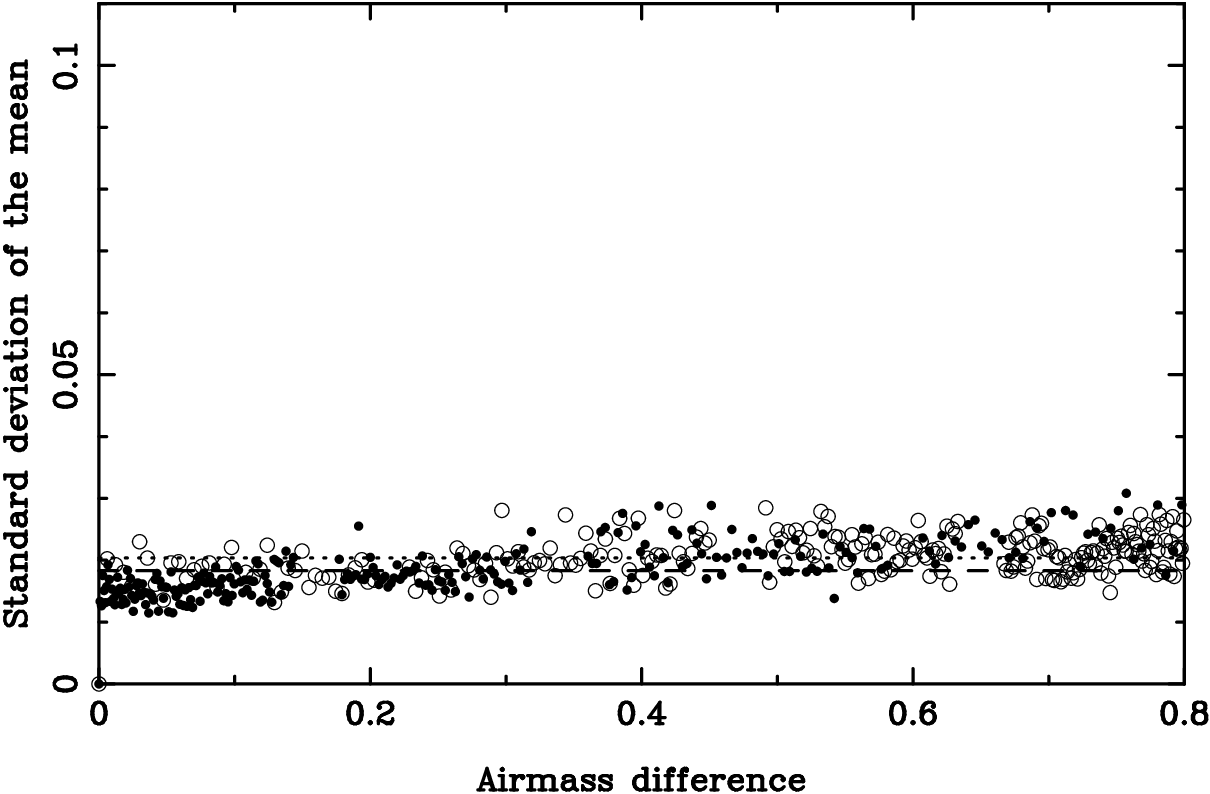}

\caption{Same as Figure \ref{twodivs} but over the range of $2.0602 -
2.0608\micron$.
\label{inlines}}

\end{figure}

We also examine the standard deviation of the spectrum as a function of
airmass difference for a region of the spectrum with small telluric
absorption, $2.0602 - 2.0608\micron$. The theoretical noise limit of 1\%
is reached (Figure \ref{inlines}) and for large airmass differences
there is only a small decrease in signal to noise. Clearly, high signal
to noise infra-red spectra are obtainable in regions free of telluric
absorption.

\subsection{Removing the residual telluric line features in oversampled
spectra}

\begin{figure}
\includegraphics[angle=270,width=\columnwidth]{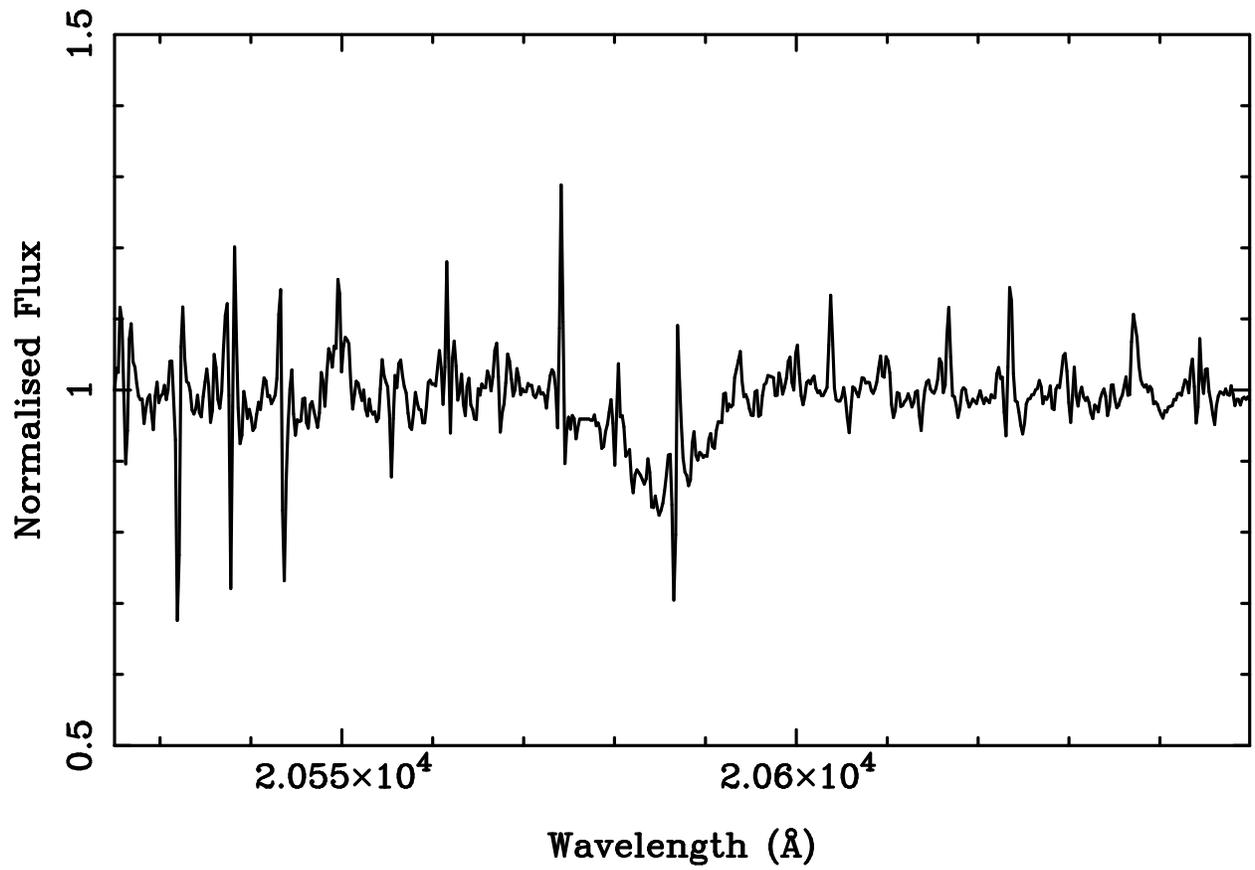}

\caption{Spectrum of HD 192639. Residuals from the division of the
standard are visible, along with the broad absorption from
\ion{He}{1} at $2.058\micron$.\label{hdclean}}

\end{figure}

Figure \ref{hdclean} shows the \ion{He}{1} absorption in a O7Ib
supergiant with data taken on 07 June 2003 UT. The science target is HD
192639 observed at 1116UT at an airmass of 1.21 and the reference star
is HD 195050 observed at 1050UT at an airmass of 1.36. The spectra are
wavelength calibrated using the telluric absorption lines. The broad
stellar absorption is clearly visible after dividing out the telluric 
lines.  The residual spikes are due to the change in airmass and the 
telluric profile structure between the two exposures.

We investigated different methods for removing the residuals, which
include simple Fourier filtering, clipping, and modeling.

We oversample our spectra above the Nyquist frequency of 2.5 pixels per
spectrally resolved element, and our measured sampling (the instrumental
profile) is 4.0 pixels. This, then, is the narrowest feature that the
spectrograph can resolve for any spectral line source. The telluric
residuals in Figure \ref{hdclean} arise from small changes in the shape
of the line profile of the telluric absorption, and subsequent division
of the science object by the standard source gives rise to large flux
changes over very small wavelength intervals. These are usually smaller
than the instrumental profile, and can be readily identified in the
final spectra. By comparing to a known list of telluric absorption lines
and to the standard star spectrum, regions of poor signal to noise
corresponding to the core of the telluric lines can be identified and
interpolated over using a low pass filter.

\begin{figure}
\includegraphics[angle=0,width=\columnwidth]{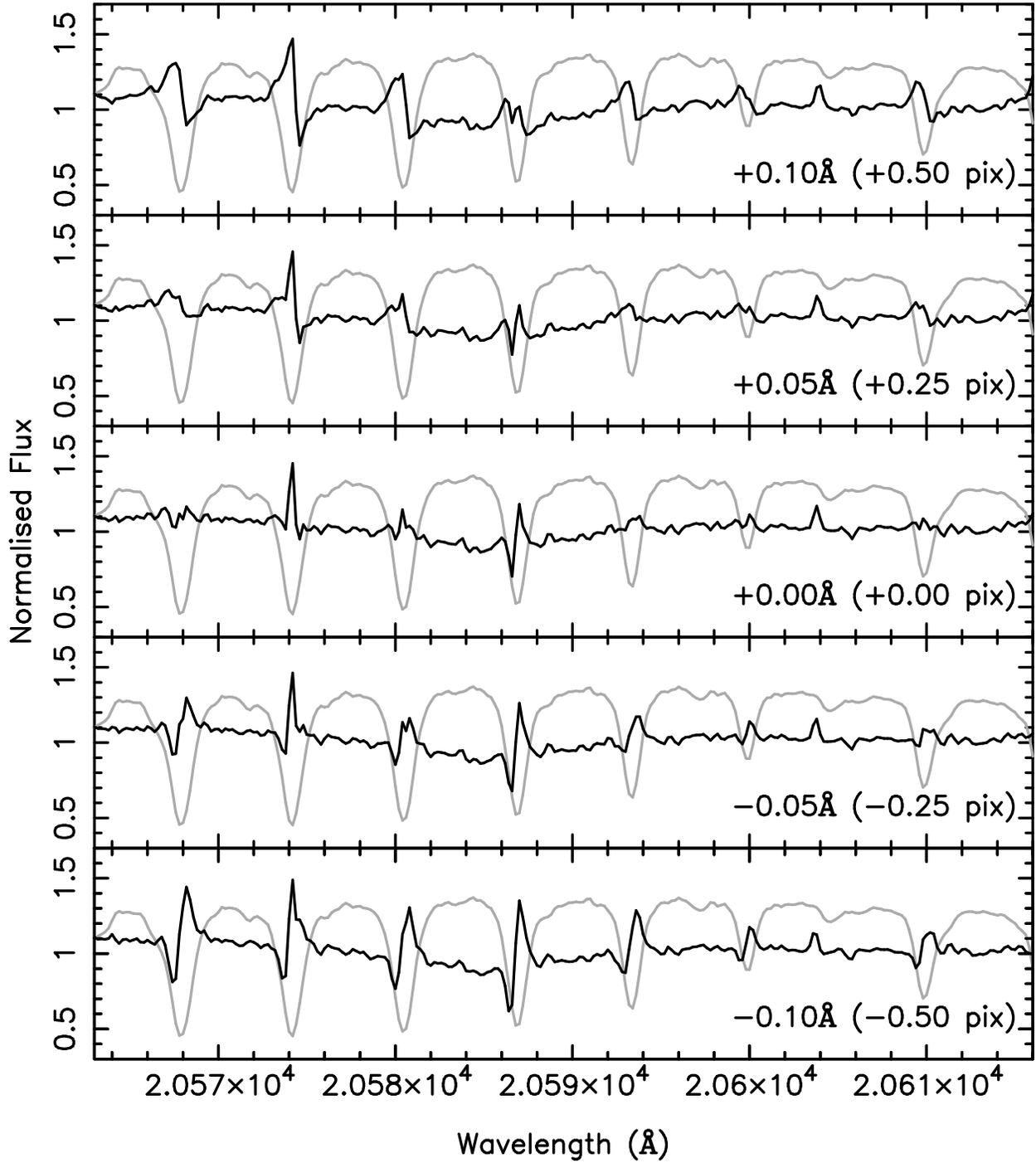}

\caption{Spectrum of HD 192639 as a function of different wavelength
misalignments. In each panel, the spectrum of HD 192639 is offset by the
amount indicated in the lower right corner before division by the
standard star spectrum. The standard star spectrum is shown in light
gray to the same scale and with a flux offset of +0.3.
\label{spectrum_real}}

\end{figure}

Another hypothesis is that the large residual spikes are due to a
systematic wavelength shift errors between to the two spectra. To study
the effect of introducing a systematic wavelength shift in one of the
two spectra, we take a detailed portion of the spectrum in Figure
\ref{hdclean} and add a wavelength offset to the science star spectrum
before dividing by the standard star. The results of adding a range of
shifts is shown in Figure \ref{spectrum_real}. Clear P Cygni-type
profiles appear for even small wavelength offsets, indicating that
accurate wavelength calibration is necessary for high spectral
resolution work of this type. On the right hand side of the spectrum,
telluric residuals of up to 40\% of the continuum remain for all
displayed shifts, whilst other telluric lines cancel out to give a flat
continuum at other wavelengths. We confirmed this by dividing the A star
spectra at airmasses similar to those of HD 192639 and its standard
star, confirming the need to match airmasses of scientific targets.

The almost constant wavelength spacing of the telluric lines in the
ro-vibrational bands suggests some form of Fourier filtering to remove
the residuals. We find that there is no effective way to remove all the
residual lines without degrading the line profile on astrophysical lines
of interest, and so we do not recommend Fourier filtering as an
effective way of removing the residuals. It is best to manually
interpolate over known telluric absorption features and then use a
piecewise spline to supply replacement values over regions of low signal
to noise.

\section{Conclusions}

The near-infrared spectral region provides a number of diagnostic
transitions important for the analysis of a variety of astronomical
objects.  Yet, ground based observations are plagued with the removal of
numerous, varying and sometimes very strong absorption features
originating in the Earth's atmosphere.  We investigate the telluric
absorption as a function of airmass.  Naturally, in the limit of bright
stars and high spectral resolution, the greater the airmass difference
between the science target and the reference star, the lower the
resultant signal to noise achieved in the final spectrum and the greater
the contamination of systematic residuals from the telluric line wings.
However, we also show that observations at higher airmasses are degraded
due to the larger rate of change of airmass with time, and with the
greater path length through the atmosphere. We find that a signal to
noise ratio of 20 (5\%) is achieveable for targets at airmasses of
around 1.9, increasing to a SNR of 50 (2\%) for observations near an
airmass of 1.15.

We conclude that high signal-to-noise (100 and greater) spectra of
astrophysical spectral lines whose wavelengths and line widths coincide
with the cores and widths of some of the deepest telluric absorption is
impossible to achieve with the observing technique described in this
paper. The variability of the telluric lines combined with the large
flux attenuation and necessarily long integration times mean that only
simultaneous observations with standards can compensate for the effects
of the atmosphere.

By using high spectral resolution spectra, regions of low absorption
between the telluric lines are able to reach near photon-limited
sensitivities (down to 1\%), and for spectral features that are
significantly broader than telluric features, high signal to noise
spectra are possible.

\acknowledgments

Thanks go to John Rayner and Paul Sears at the IRTF, and to Tom Greene
and Tom Geballe for their suggestions and additional insight. We also
thank our anonymous referees for their comments and corrections. This
work is supported by the National Science Foundation under grant AST
00-94050 to the University of Cincinnati.

\clearpage

\clearpage

\clearpage

\clearpage

\clearpage

\clearpage

\clearpage

\end{document}